\documentclass{statsoc}

\usepackage{graphicx}
\usepackage{amssymb}

\title[Penalized Orthogonal-Components Regression]{Penalized Orthogonal-Components Regression for Large $p$ Small $n$ Data}%

\author[Zhang {\it et al.}]{Dabao Zhang, Yanzhu Lin and Min Zhang}
\address{Department of Statistics, Purdue University, West Lafayette, IN, USA.}%

\begin{document}

\begin{abstract}
We propose a penalized orthogonal-components regression (POCRE) for large $p$ small $n$ data. Orthogonal components are sequentially constructed to maximize, upon standardization, their correlation to the response residuals. A new penalization framework, implemented via empirical Bayes thresholding, is presented to effectively identify sparse predictors of each component. POCRE is computationally efficient owing to its sequential construction of leading sparse principal components. In addition, such construction offers other properties such as grouping highly correlated predictors and allowing for collinear or nearly collinear predictors. With multivariate responses, POCRE can construct common components and thus build up latent-variable models for large $p$ small $n$ data.


\end{abstract}

\section{Introduction}


Available high-throughput biotechnologies make it possible to comprehensively analyze genomic, proteomic, or metabolomic profiles of biological samples, thus identifying molecular signatures to understand complex biological systems. Such profile analysis holds an enormous promise for its use in early disease detection, assessment of prognosis, measurement of drug efficacy, and eventually, personalized medicine. However, it usually entails collection of a massive amount of possible predictors (i.e., {\it large $p$}) from each of a small number of biological individuals (i.e., {\it small $n$}), and therefore identifying the underlying sparse predictors presents a task of ``finding a very few needles in a haystack". The structured and noisy predictors make the task even more difficult.


\cite{Breiman96} showed that classical step-wise regression is unstable since modifying a single observation can change the fitted model significantly. On the other hand, ridge regression is stable but it lacks the ability to select variables. \cite{Tibshirani96} employed an $\ell_1$-norm penalty and proposed the lasso method, which gained popularity due to its ability to select variables and, at the same time, exhibit the stability of ridge regression. This method has a Bayesian interpretation with independent Laplace priors (\cite{Tibshirani96}; \cite{Park08}). However, lasso lacks the grouping property, that is, it tends to select one predictor from a group of highly correlated predictors, see \cite{Zou05} for more details.


The grouping property plays an important role in analyzing $p\gg n$ data with clustered but noisy predictors. The predictors for molecular signatures are naturally grouped due to sharing metabolomic pathways or biological processes, and are preferred to be included or excluded from the model simultaneously. On the other hand, highly correlated predictors can borrow strength from each other to counter the noise effect. Many lasso variants have therefore been proposed to take advantage of the grouped predictors either implicitly or explicitly. For example, \cite{Zou05} proposed the elastic net (EN) which added a $\ell_2$-norm penalty; \cite{Tibshirani05} proposed the fused lasso including another $\ell_1$-norm penalty to encourage similarity between coefficients; and \cite{Yuan06} proposed the group lasso which modified the $\ell_1$-norm penalty for grouped coefficients.



Another strategy in analyzing $p\gg n$ data is to first reduce the dimension of predictors by constructing components, i.e., ``eigen" predictors, and then fit regression models by applying step-wise approaches to these components. Such construction of components not only provides a potential solution to the ``curse of dimensionality", but also groups predictors which are highly correlated or share certain common coherent patterns. Both unsupervised and supervised dimension reduction methods have been proposed. While many unsupervised methods have been proposed on the basis of principal component analysis (PCA; \cite{Hastie00}, \cite{Bair06}, \cite{Cook07}), the partial least squares (PLS; \cite{Garthwaite94}) regression is a supervised approach and has been widely used in chemometrics and bioinformatics, see \cite{Kramer98}, and \cite{Nguyen02}, among others. 

In this paper we propose a penalized orthogonal-components regression (POCRE) via a new penalization framework which can effectively identify sparse predictors from a large number of candidates. Section 2 presents the general idea of orthogonal-components regression, and the penalized orthogonal-components regression is proposed in Section 3. The penalization is implemented in Section 4 using the empirical Bayes thresholding proposed by \cite{Johnstone04}. Such implementation allows adaptively identifying sparse predictors and leads to the computationally efficient POCRE algorithm which is summarized in Section 5. Simulation studies and real data analysis are shown in Section 6 and 7 respectively. We conclude this paper with a discussion.



\section{Orthogonal-Components Regression}


To illustrate the ideas behind the orthogonal-components regression, we assume
\begin{eqnarray} \label{Eqn-ClassicReg}
Y= \beta^T X + \epsilon,
\end{eqnarray}
where $Y$ is a $k$-dimensional column vector, $X$ is a $p$-dimensional column vector independent of $\epsilon$, $E[X] = 0$, and $\beta$ is a $p\times k$ matrix. When $var(X)$ is non-singular and the sample size $n$ is reasonably larger than $p$, either likelihood method or moment method can provide a satisfactory estimate of $\beta$.

Here we are interested in estimating $\beta$ in the large $p$ paradigm. First, $var(X)$ may be singular or nearly singular due to collinear or highly correlated predictors in $X$. Second, when $p$ is too large, it is usually infeasible to assume that the sample size $n$ is larger than $p$. In either case, it is difficult, if not impossible, to estimate $\beta$ using the classical methods.

To avoid possible problems with large $p$, we construct orthogonal components as linear combinations of all predictors in $X$, and then regress $Y$ on these orthogonal components. Such orthogonal components can be sequentially constructed. Specifically, let $\tilde{X}_1 = X$ and $\tilde{Y}_1 = Y$. The first component $\omega_1^T \tilde{X}_1$ is constructed with $\omega=\omega_1$ maximizing $\|cov(\tilde{Y}_1,\omega^T\tilde{X}_1)\|^2$ under the condition $\|\omega\|=1$. Since
\[
\|cov(\tilde{Y}_1,\omega^T \tilde{X}_1)\|^2 = \|cov(Y,\omega^T X)\|^2,
\]
$\omega_1$ is the leading eigenvector of $cov(Y,X)^T cov(Y,X)$. Here the leading eigenvector refers to the one with the largest eigenvalue. When $Y$ is univariate, i.e., $k=1$, $\omega_1 \propto cov(Y,X)^T$.

After constructing the $j$-th component $\omega_j^T \tilde{X}_j$, we then remove $\omega_j^T \tilde{X}_j$ from $\tilde{X}_j$ such that $\tilde{X}_{j+1} = \tilde{X}_j - \theta_{j} \omega_j^T \tilde{X}_j$ is uncorrelated to $\omega_j^T \tilde{X}_j$, i.e.,
\[
cov(\tilde{X}_{j+1}, \omega_j^T \tilde{X}_j) = 0 \Longrightarrow \theta_{j} = \frac{var(\tilde{X}_j)\omega_j}{\omega_j^T var(\tilde{X}_j) \omega_j}.
\]
We also remove $\omega_j^T \tilde{X}_j$ from $\tilde{Y}_j$ such that $\tilde{Y}_{j+1} = \tilde{Y}_j - \vartheta_{j} \omega_j^T \tilde{X}_j$ is uncorrelated to $\omega_j^T \tilde{X}_j$, i.e.,
\[
cov(\tilde{Y}_{j+1}, \omega_j^T \tilde{X}_j) = 0 \Longrightarrow \vartheta_{j} = \frac{cov(\tilde{Y}_j,\tilde{X}_j)\omega_j}{\omega_j^T var(\tilde{X}_j) \omega_j}.
\]
Then the $(j+1)$-st component $\omega_{j+1}^T \tilde{X}_{j+1}$ is constructed with $\omega = \omega_{j+1}$ maximizing
\begin{eqnarray*} 
\|cov(Y,\omega^T \tilde{X}_{j+1})\|^2 = \|cov(\tilde{Y}_{j+1},\omega^T \tilde{X}_{j+1})\|^2
\end{eqnarray*}
under the condition $\|\omega\|=1$. Note that $\omega_{j+1}$ is the leading eigenvector of $cov(Y,\tilde{X}_{j+1})^T \times cov(Y,\tilde{X}_{j+1})$. When $k=1$, $\omega_{j+1}$ equals to the normalized $cov(Y,\tilde{X}_{j+1})^T$.

This construction stops whenever $Y$ is uncorrelated to $\tilde{X}_j$. Since
\[
\omega_j^T\tilde{X}_j = \omega_j^T(I-\theta_{j-1}\omega_{j-1}^T) \tilde{X}_{j-1} = \cdots = \omega_j^T \left\{\prod_{l=1}^{j-1} (I-\theta_{j-l}\omega_{j-l}^T)\right\} X,
\]
we denote the $j$-th component as $\varpi_j^T X$. Upon the completion of the construction, $\varpi_1^T X$, $\varpi_2^T X$, $\cdots$, are uncorrelated, i.e., they constitute a sequence of orthogonal components, which lead to the orthogonal-components regression model.

\vskip6pt
\noindent{\it Theorem 1.} $\varpi_1^T X$, $\varpi_2^T X$, $\cdots$, are orthogonal, i.e., uncorrelated. Furthermore,
\begin{eqnarray}\label{Eqn-OCReg}
E[Y|X] = \sum_j \vartheta_j \left(\varpi_j^T X\right).
\end{eqnarray}
\vskip6pt


Compared to the original regression (\ref{Eqn-ClassicReg}), the orthogonal-components regression (\ref{Eqn-OCReg}) can be fit by only calculating the eigenvectors of matrices but not the inverses, which makes it appealing in analyzing $p\gg n$ data. Furthermore, if the predictors are highly correlated or even collinear, the orthogonal-components regression is still able to provide robust solution. The calculation is very fast due to the fact that $\varpi_1^T X$, $\varpi_2^T X$, $\cdots$, can be easily constructed and that they are uncorrelated.

\section{Penalized Orthogonal-Components Regression}

Implementing the orthogonal-components regression (\ref{Eqn-OCReg}) is subject to finding the leading eigenvector of $cov(Y,\tilde{X}_j)^T cov(Y,\tilde{X}_j)$ to construct the $j$-th component $\varpi_j^T X$. However, the involved covariances are not observed and need to be estimated from the observed data, say the i.i.d. sample $(\mathbf{Y}_{n\times k},\mathbf{X}_{n\times p})$. \cite{Wold75} estimated the covariances with their empirical estimates and proposed the partial least squares. Each subsequently constructed component is a linear combination of all available predictors. In the case of $p\gg n$ data, especially when only a small number of predictors contribute to the response variables, the results from partial least squares regression inflate the errors besides the difficulty in interpreting the results. Here we will pursue a penalized construction for sparse loadings.

Let
\[
\mathbf{M} = \widehat{cov}(Y,\tilde{X}_j),
\]
be an estimate of $cov(Y,\tilde{X}_j)$. A major step in implementing the orthogonal-components regression is to find the leading sparse eigenvector of $\mathbf{M}^T\mathbf{M}$. The following theorem by \cite{Zou06} implies that finding the leading eigenvector can be taken as an optimization problem, which sheds light on constructing sparse eigenvectors.

\vskip6pt
\noindent{\it Theorem 2. (\cite{Zou06})} For any $\kappa>0$, let
\begin{eqnarray} \label{Eqn-PCA}
(\tilde{\alpha},\tilde{\gamma}) = argmin_{\alpha,\gamma: \|\alpha\|=1} \left\{ \|\mathbf{M}-\mathbf{M}\gamma\alpha^T\|^2 + \kappa \|\gamma\|^2 \right\}.
\end{eqnarray}
Then, $\omega = \tilde{\gamma}/\|\tilde{\gamma}\|$ is the leading eigenvector of $\mathbf{M}^T\mathbf{M}$, i.e., $\mathbf{M}^T\mathbf{M}\omega = c\omega$ where $c$ is the largest eigenvalue of $\mathbf{M}^T\mathbf{M}$.
\vskip6pt

To ensure a sparse principal component, we consider a general version of the criterion (\ref{Eqn-PCA}), i.e., with tuning parameter $\lambda$ and penalty function $p_{\lambda}(\gamma)$,
\begin{eqnarray} \label{Eqn-PPCA}
(\hat{\alpha}(\kappa),\hat{\gamma}(\kappa)) = argmin_{\alpha,\gamma: \|\alpha\|=1} \left\{ \|\mathbf{M}-\mathbf{M}\gamma\alpha^T\|^2 + \kappa \|\gamma\|^2 + p_{\lambda}(\gamma) \right\}.
\end{eqnarray}
Here the penalty is introduced to benefit estimating covariances and thresholding $\gamma$ such that most of the elements in $\gamma$ are zero, i.e., $\gamma$ is sparse. While Theorem 2 implies that specific value of $\kappa$ does not affect the solution to optimization problem (\ref{Eqn-PCA}), the following theorem states that sparse $\gamma$ can be derived from a problem without specifying $\kappa$ in (\ref{Eqn-PPCA}).

\vskip6pt
\noindent{\it Theorem 3.} Suppose $p_{\lambda}(c \gamma) = c p_{\lambda}(\gamma)$ for any scaler $c>0$. Let $(\hat{\alpha}(\kappa), \hat{\gamma}(\kappa))$ be the solution to (\ref{Eqn-PPCA}). And $(\hat{\alpha}, \hat{\gamma})$ is the solution to the following problem
\begin{eqnarray} \label{Eqn-FPPCA}
(\hat{\alpha},\hat{\gamma}) = argmin_{\alpha,\gamma: \|\alpha\|=1} \left\{ -2 \gamma^T\mathbf{M}^T\mathbf{M}\alpha + \|\gamma\|^2 + p_{\lambda}(\gamma) \right\}.
\end{eqnarray}
Then, $\hat{\gamma}(\kappa)/\|\hat{\gamma}(\kappa)\|$ approaches to $\hat{\gamma}/\|\hat{\gamma}\|$ when $\kappa \rightarrow \infty$.
\vskip6pt

We will iteratively solve (\ref{Eqn-FPPCA}) for $\hat{\alpha}$ and $\hat{\gamma}$. First, for a given $\gamma$, we have
\[
\hat{\alpha}(\gamma) = argmin_{\alpha: \|\alpha\|=1} \left\{ -2 \gamma^T\mathbf{M}^T\mathbf{M}\alpha \right\} = \mathbf{M}^T\mathbf{M}\gamma/\|\mathbf{M}^T\mathbf{M}\gamma\|.
\]
Second, for a given $\alpha$, we have
\begin{eqnarray} \label{Eqn-Gamma}
\hat{\gamma}(\alpha) = argmin_{\gamma} \left\{ \|\gamma - \mathbf{M}^T\mathbf{M}\alpha\|^2 + p_{\lambda}(\gamma) \right\},
\end{eqnarray}
which will be approximated using the empirical Bayes thresholding as discussed in the following section.

\section{Penalization via Empirical Bayes Thresholding}

Denote $\mathbf{Z}= \mathbf{M}^T\mathbf{M} \alpha$. Then solving for $\hat{\gamma}(\alpha)$ in (\ref{Eqn-Gamma}) is subject to minimizing $\|\mathbf{Z}-\gamma\|^2 + p_{\lambda}(\gamma)$ with respect to $\gamma$. Suppose the $i$-th component of $\mathbf{Z}$ is $z_i$, and further assume,
\[
z_i = \mu_i + \epsilon_i,\ \ \ \epsilon_i \sim N(0,\sigma^2).
\]
Since $p$ is large and most of $\{\mu_i, 1\le i\le p\}$ are zero, the variance $\sigma^2$ can be estimated by
\begin{eqnarray}\label{Eqn-ESigma}
\hat{\sigma} = median_{1\le i\le p}\left\{ |z_i| \right\}/\Phi^{-1}(0.75).
\end{eqnarray}
Note that this estimate partially accounts for under- or over-dispersion due to dependent data, see~\cite{Efron04a}. When implementing the penalization of POCRE, we also introduce a tuning parameter $\lambda$ to account for the possible over-dispersion when standardizing $z_i$ using $\lambda \hat{\sigma}$. Without loss of generality, hereafter we assume $\epsilon_i \stackrel{iid}{\sim} N(0,1)$.

When $p_{\lambda}(\cdot)$ is specified by the logarithm of a prior density function, the optimal $\gamma$ is indeed a Bayesian estimate of $(\mu_1, \cdots, \mu_p)^T$. In consideration of the sparsity of $\gamma$, we employ the empirical Bayes thresholding (EBT) proposed by \cite{Johnstone04, Johnstone05} for a better approximation to the leading sparse eigenvalue of $\mathbf{M}^T\mathbf{M}$.

Specifically, we assume a mixture prior with a point mass at zero and a quasi-Cauchy distribution for each $\mu_i$, i.e.,
\[
\pi(\mu) = (1-w) \delta_{0}(\mu) + w \frac{1}{\sqrt{2\pi}}\left\{ 1-\frac{|\mu_i|\Phi(-|\mu_i|)}{\phi(\mu_i)}  \right\},
\]
where $\delta_0(\cdot)$ is Dirac's delta function. Since the marginal distribution of $z_i$ is
\[
g(z_i) = \frac{1-w}{\sqrt{2\pi}}e^{-z_i^2/2} + \frac{w}{\sqrt{2\pi}z_i^2} \left(1-e^{-z_i^2/2}\right),
\]
an estimate of $w$, say $\hat{w}$, can be calculated by maximizing the marginal likelihood. Then $\mu_i$ can be estimated by the posterior median, i.e.,
\[
\hat{\mu}_i = \hat{\mu}(z_i) =  median(\mu_i|z_i, \hat{w}).
\]
As $\hat{w}$ provides a data-driven estimate of the parameter sparsity, the resultant estimate is adaptive to the sparsity of the underlying parameter. \cite{Johnstone04} also showed that the empirical Bayes estimator $\hat{\mu}(z)$ is a thresholding estimator in the sense that (i) $\hat{\mu}(z)$ is increasing on $z\in R$; (ii) $|\hat{\mu}(z)|\le |z|$, $\forall z\in R$; (iii) $\hat{\mu}(-z)=-\hat{\mu}(z)$; (iv) there exists $\tau>0$ such that $\hat{\mu}(z)=0$ if and only if $|z|\le \tau$.

As noted above, although $\hat{\mu}_i$ is constructed by assuming all components of $\mathbf{Z}$ are independent, using the estimate $\hat{\sigma}$ in (\ref{Eqn-ESigma}) and the tuning parameter $\lambda$ in the penalty function $p_{\lambda}(\cdot)$ account for possible dependence. In practice, ten-fold cross-validation can be employed to elicit the optimal value of $\lambda$ ranging from $0.6$ to 1. As demonstrated by our simulation studies, it usually suffices to consider $\lambda\in\{0.8, 0.81, 0.82, \cdots, 1\}$.




\section{The Algorithm}

Without loss of generality, we further assume that both $\mathbf{X}$ and $\mathbf{Y}$ are centered. Therefore, an estimate of $cov(Y,X)$ is $\mathbf{M}\propto \mathbf{Y}^T\mathbf{X}$. Suppose $\omega_1, \cdots, \omega_{j-1}$ have been calculated, and $\mathbf{X}_j$ has been updated accordingly. An estimate of $cov(Y,\tilde{X}_j)$ is proportional to $\mathbf{Y}^T\mathbf{X}_j$. We can therefore proceed to find $\omega_{j}$ as follows,
\begin{description}\itemsep=0pt
\item[]{\bf 1.} Initialize $\gamma$ to be the leading eigenvector of $\mathbf{X}_j^T \mathbf{Y} \mathbf{Y}^T \mathbf{X}_j$;
\item[]{\bf 2.} Update $\alpha = \mathbf{X}_j^T\mathbf{Y}\mathbf{Y}^T\mathbf{X}_j\gamma/\|\mathbf{X}_j^T\mathbf{Y}\mathbf{Y}^T\mathbf{X}_j\gamma\|$;
\item[]{\bf 3.} Calculate $\hat{\sigma} = median\left\{ |\mathbf{X}_j^T\mathbf{Y}\mathbf{Y}^T\mathbf{X}_j \alpha| \right\}/\Phi^{-1}(0.75)$;
\item[]{\bf 4.} Update $\gamma = \hat{\mu}\left(\frac{\mathbf{X}_j^T\mathbf{Y}\mathbf{Y}^T\mathbf{X}_j \alpha}{\lambda\hat{\sigma}}\right) \lambda\hat{\sigma}$;
\item[]{\bf 5.} Repeat 2 -- 4 until convergence, then $\omega_{j}=\gamma/\|\gamma\|$;
\item[]{\bf 6.} Calculate $\eta_{j}=\mathbf{X}_j \omega_{j}$;
\item[]{\bf 7.} Calculate $P_{j}=\eta_{j}^T \mathbf{X}_j/\eta_j^T \eta_{j}$, and update $\mathbf{X}_{j+1} = \mathbf{X}_j-\eta_{j} P_j$.
\end{description}

Note that the first five steps are used to calculate the first principal component of
$\mathbf{X}_j^T\mathbf{Y}\mathbf{Y}^T\mathbf{X}_j$, which is adaptive to the sparsity of the non-zero loadings. Among these steps, the first step may be easily implemented using the following power method (\cite{Stewart74}), which has been used for the nonlinear iterative partial least squares (NIPALS; \cite{Wold75}),
\begin{description} \itemsep=0pt
\item[]{\bf 1.a.} Initialize $\psi$ to be the first column of $\mathbf{Y}_{j}$;
\item[]{\bf 1.b.} $\gamma = \mathbf{X}_{j}^T \psi/\|\mathbf{X}_{j}^T \psi\|$;
\item[]{\bf 1.c.} $\eta = \mathbf{X}_j \gamma$;
\item[]{\bf 1.d.} $\varphi = \mathbf{Y}^T \eta/\|\mathbf{Y} \eta\|$;
\item[]{\bf 1.e.} $\psi = \mathbf{Y} \varphi$;
\item[]{\bf 1.f.} Repeat 1.b -- 1.e until the convergence of $\gamma$.
\end{description}

When $\omega_{j}$ converges to the leading eigenvector of $\mathbf{X}_j^T\mathbf{Y}\mathbf{Y}^T\mathbf{X}_j$, then $\eta_{j}$ is an eigenvector of $\mathbf{X}_j \mathbf{X}_j^T \mathbf{Y} \mathbf{Y}^T$, which defines the $j$-th orthogonal component. Note that $P_j$ in Step 7 helps calculate $\mathbf{X}_{j+1}$ due to the fact that $\eta_j^T \mathbf{X}_{j+1} = 0$. 

Since
\[
\mathbf{X}_{j+1} = \mathbf{X}_j-\eta_{j} P_{j} = \mathbf{X}_j(I-\omega_j P_j),
\]
when writing $\mathbf{X}_{j+1} = \mathbf{X} \zeta_{j+1}$, $\zeta_{j+1}$ can be sequentially calculated as follows,
\[
\zeta_1 = I_{p\times p}; \ \ \ \zeta_{j+1} = \zeta_{j}(I-\omega_j P_j),\ j=1, 2, \cdots.
\]

Suppose that the above algorithm stops at $(l+1)$-st step, i.e., $\omega_{l+1}=0$. Then we regress $\mathbf{Y}$ on the orthogonal components $\eta_j$, $j=1, 2, \cdots, l$, and fit the following model,
\[
\hat{\mathbf{Y}} = \sum_{j=1}^l \eta_j Q_j,
\]
which implies that $Q_{j}=\eta_{j}^T \mathbf{Y} /\eta_j^T \eta_{j}$. Since
$\eta_j = X\zeta_j\omega_j$, the estimate $\hat{\beta}$ of $\beta$ in (\ref{Eqn-ClassicReg}) can then be derived as
\[
\hat{\beta} = \sum_{j=1}^l \zeta_j\omega_j Q_j.
\]

\section{Simulation Studies}

We consider five different cases of large $p$ small $n$ data to evaluate the performance of POCRE and compare with other approaches such as partial least squares (PLS), ridge regression, lasso, and elastic net (EN). The first two cases have highly and mildly correlated predictors respectively, the third one has clustered predictors, the fourth one demonstrates a measurement-error model, and the fifth one features a latent-variable model. In all cases, we fix $p=1000$ and consider both $n=50$ and $n=100$.

\vskip6pt
{\bf Case 1 (High Correlations).} $Y=2\sum_{j=1}^{10} X_{j}+\sum_{j=101}^{110} X_{j}+\varepsilon$, where $\varepsilon \sim N(0,1)$, and each block $\{X_{k+1},\cdots,X_{k+100}\}$ is simulated from an AR(1) process with $\rho=0.9$, $k=0, 100, \cdots, 900$.

\vskip6pt

{\bf Case 2 (Mild Correlations).} Same as Case 1 except that $\rho=0.5$.

\vskip6pt

{\bf Case 3 (Clustered Predictors).} $Y=1.5 \sum_{j=1}^{30} X_{j}+\varepsilon$, where $\varepsilon \sim N(0,15^{2})$, and $X_j = Z_1 1_{\{j\le 10\}}+Z_2 1_{\{11\le j\le 20\}} + Z_3 1_{\{21\le j\le 30\}}+\xi_j$. Here $Z_{1}, Z_{2},Z_{3} \stackrel{iid}{\sim} N(0,1)$, and $\xi_j \stackrel{iid}{\sim} N(0,0.01)$.

\vskip6pt

{\bf Case 4 (Errors in Predictors).} $Y=Z_{1}+2Z_{2}+Z_{3}+\varepsilon$, where $\varepsilon \sim N(0,1)$. Note that $X_j = sign(5.5-j) Z_1 1_{\{j\le 10\}} + sign(15.5-j)Z_2 1_{\{11\le j\le 20\}} + Z_3 1_{\{21\le j\le 30\}}+\xi_j$, where $Z_{1}, Z_{2},Z_{3} \stackrel{iid}{\sim} N(0,1)$, and $\xi_j \stackrel{iid}{\sim} N(0,1)$.

\vskip6pt

{\bf Case 5 (Latent-Variable Model).} $Y_k = a_k Z_1 + b_k Z_2 + \varepsilon_k$, $1\le k\le 5$, where $a_1=a_2=b_2=2$, $b_1=a_3=b_3=-2$, $a_4=a_5=3$, $b_4=-b_5=1.5$, and $\varepsilon_k \stackrel{iid}{\sim} N(0,1)$. $Z_1 = X_{50}+X_{150}+X_{250}+X_{350}+X_{450}+X_{550}$ and $Z_2 = X_{51}+X_{153}+X_{256}+X_{359}+X_{467}+X_{583}$, where $X$'s are the same as in Case 1 except that $\rho=0.3$.


\vskip6pt

Here we evaluate the algorithms on the basis of two different criteria, i.e.,  the loss defined as $E[\|Y-\hat Y\|^2\big|\hat{\beta}]-tr\{var(Y|X)\}$, and the false discovery rate (FDR). In each case, we simulated $100$ datasets, and therefore calculated the values of the loss and FDR on the basis of the estimated parameters. Ten-fold cross-validations are used to find the optimal tuning parameters for EN, lasso, POCRE, and ridge regression, and the optimal number of components for PLS.

Since neither PLS nor ridge regression selects variables and both instead build up the model using all available predictors, FDR is not reported for either method. In all cases, both methods report very large losses compared to the other three methods due to inflated prediction errors by using all predictors. It is interesting to note that both PLS and ridge regression perform similarly in terms of losses, although PLS is able to build common components for multivariate responses.

In Case 1 with highly correlated predictors, both lasso and POCRE present much smaller losses than EN, as shown in Table~\ref{Table-LOSS}. When the correlations between predictors are mild as in Case 2, the losses of both EN and POCRE dramatically decrease but the loss of lasso increases when $n=100$. For $n=50$, all three methods increase the losses with lasso increases the most. In both cases, lasso presents the smallest losses. However, POCRE is able to build up common components shared by multiple responses and lowers the losses, as shown in Case 5. Indeed, POCRE has much smaller loss than other methods for $n=100$, and is comparable to lasso for $n=50$.

\begin{table}
\caption{\label{Table-LOSS} Summary on losses (with standard errors in parentheses)}
\centering
\begin{tabular}{c | c c c c c c}
\hline\hline
$n$ & Method & Case 1 & Case 2 & Case 3 & Case 4 & Case 5 \\ \hline
& EN & 29.80(1.31) & 2.03(1.53) & 103.34(4.35) & 1.45(0.04) & 13.48(1.29) \\
& Lasso & {\bf 0.66}(0.02) & {\bf 1.76}(0.10) & 72.12(4.04) & 1.59(0.03) & 12.47(0.79) \\
100 & PLS & 81.44(1.15) & 89.94(0.48) & 187.57(3.25) & 3.10(0.02) & 254.43(0.79) \\
& POCRE & 6.13(0.53) & 3.58(0.42) & {\bf 14.93}(2.81) & {\bf 0.87}(0.03) & {\bf 4.74}(1.99) \\
& Ridge & 81.60(1.13) & 89.71(0.44) & 193.90(3.21) & 3.09(0.02) & 253.18(0.52) \\
\hline
& EN & 39.23(2.09) & 52.45(2.65) & 141.90(7.93) & 2.30(0.13) & 250.51(2.92)\\
& Lasso & {\bf 1.98}(0.13) & {\bf 33.24}(1.66) & 167.93(9.64) & 2.74(0.06) & {\bf 234.97}(3.21) \\
50 & PLS & 196.82(2.25) & 111.26(0.73) & 331.31(4.35) & 4.24(0.03) & 273.23(0.83)\\
& POCRE & 9.10(2.00) & 40.88(2.05) & {\bf 62.69}(5.78) & {\bf 1.78}(0.06) & 236.53(5.17) \\
& Ridge & 192.01(2.26) & 110.56(0.53) & 333.79(4.45) & 4.22(0.03) & 269.71(0.62) \\
\hline
\end{tabular}
\end{table}

In Case 3 with clustered predictors, POCRE performs extremely well when compared to all other methods. In Case 4 with errors in predictors, POCRE also presents the smallest losses. Indeed, in Case 3, POCRE decreases $55.82\%$ and $79.30\%$ of the losses when compared to the best of all other methods for $n=50$ and $n=100$, respectively. And in Case 4, POCRE decreases $22.61\%$ and $40.00\%$ for $n=50$ and $n=100$, respectively. Therefore, POCRE prevails in handling clustered or noisy predictors due to its building up components through maximizing their correlations to the response variables.

In all cases, POCRE performs the best in terms of FDR, as shown in Table~\ref{Table-Count}. With $n=100$, POCRE can control the FDR under $25\%$ for all cases except Case 1 in which the FDR is at $57.45\%$ as POCRE tends to include predictors which are highly correlated to those true predictors. On the other hand, lasso presents FDR as high as $84.21\%$, with the lowest level at $57.45\%$. Not surprisingly, EN performs better than lasso in Case 3, i.e., with the lowest FDR at $41.18\%$, as it can account for group effects of predictors. However, it presents higher FDRs than lasso for all other cases. With $n=50$, although POCRE still presents lower FDRs than other two methods, all methods present high FDRs except that POCRE has the FDR at $18.92\%$ in Case 3.

\begin{table}
\caption{\label{Table-Count} Summary on FDR}
\centering
\begin{tabular}{c | c c c c c c}
\hline\hline
$n$ & Method & Case 1 & Case 2 & Case 3 & Case 4 & Case 5
\\ \hline
 & EN & 0.9603 & 0.7260 & 0.4118 & 0.7216 & 0.8452 \\
100 & Lasso & {\bf 0.5745} & 0.7037 & 0.7931 & 0.6087 & 0.8421 \\
 & POCRE & {\bf 0.5745} & {\bf 0.1304} & {\bf 0.0909} & {\bf 0.1724} & {\bf 0.2500} \\
\hline
 & EN & 0.9184 & 0.8365 & 0.7285 & 0.8167 & 0.9622 \\
50 & Lasso & 0.4722 & 0.6818 & 0.8222 & 0.6333 & 0.8197 \\
 & POCRE & {\bf 0.3103} & {\bf 0.5102} & {\bf 0.1892} & {\bf 0.4194} & {\bf 0.7742} \\
\hline\hline
\end{tabular}
\end{table}

\section{A Real Data Analysis}

\cite{Lan06} designed an experiment to identify the genetic basis for differences between two inbred mouse populations (B6 and BTBR). A total of 60 arrays were used to monitor the expression levels of 22,690 genes of 31 female and 29 male mice. Some physiological phenotypes, including numbers of stearoyl-CoA desaturase 1 (SCD1), glycerol-3-phosphate acyltransferase (GPAT) and phosphoenopyruvate carboxykinase (PEPCK), were also measured by quantitative real-time RT-PCR. The gene expression data and the phenotypic data are available in GEO (http://www.ncbi.nlm.nih.gov/geo; accession number GSE3330).

We adjusted the phenotypic values to remove the possible gender effects. For each phenotype, its correlation to each gene is calculated, then an overall correlation coefficient (OCC) of the three phenotypes to a single gene is defined as minimizing the absolute values of the correlation coefficients between the gene and three phenotypes. Here we investigated expression profiling of the top 5,000 genes (ranked on the basis of OCC) to predict the three physiological phenotypic values. We set up the test dataset including randomly selected 5 female and 5 male mice, and the rest are included in the training dataset. We built up the model using the training dataset and then calculated the sum of squared prediction errors (SSPE) using the test data.

\begin{figure}

\begin{minipage}[h]{0.32\linewidth}
\centering
\makebox{\includegraphics[scale=0.32]{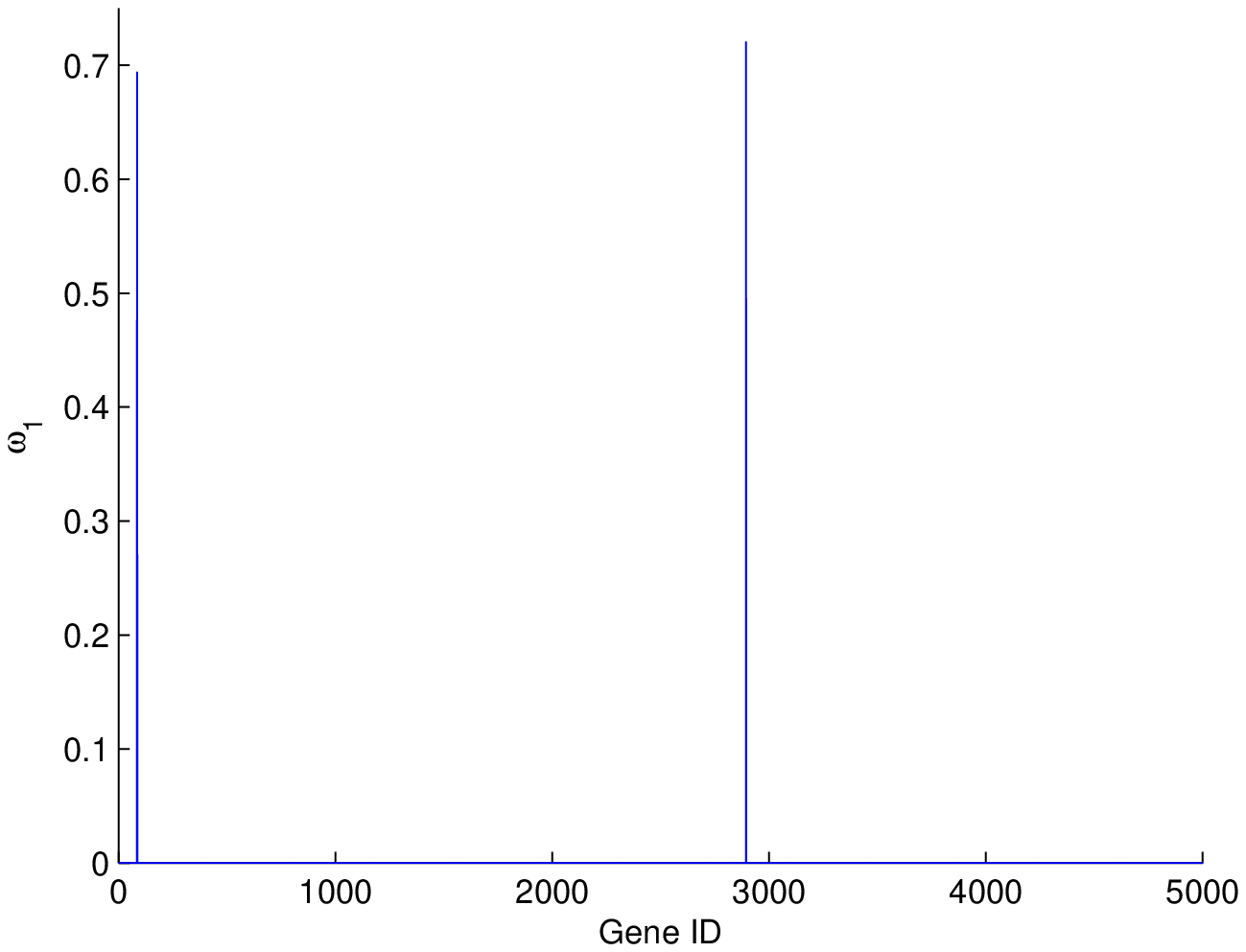}}
\end{minipage}
\begin{minipage}[h]{0.32\linewidth}
\centering
\makebox{\includegraphics[scale=0.32]{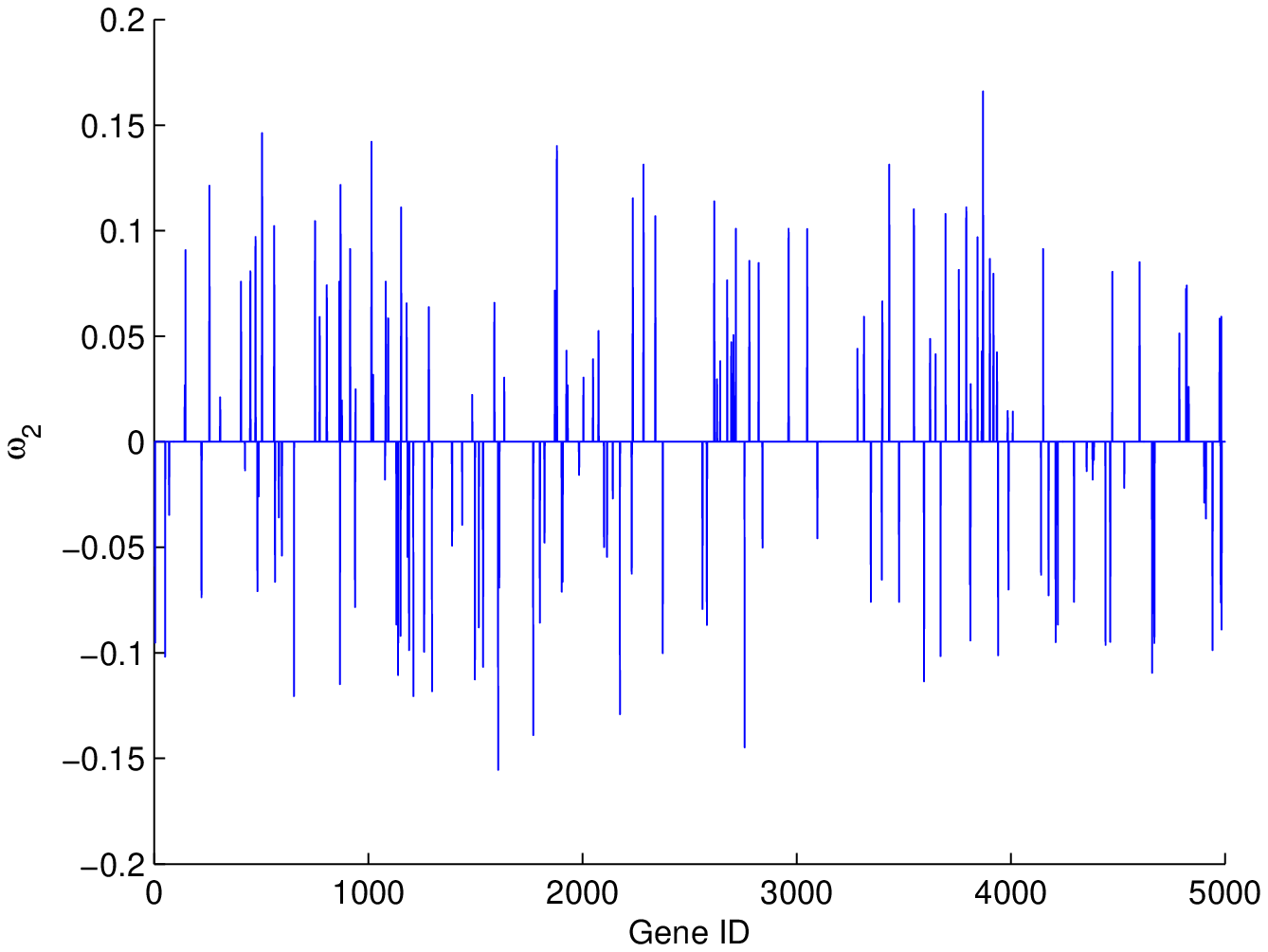}}
\end{minipage}
\begin{minipage}[h]{0.32\linewidth}
\centering
\makebox{\includegraphics[scale=0.32]{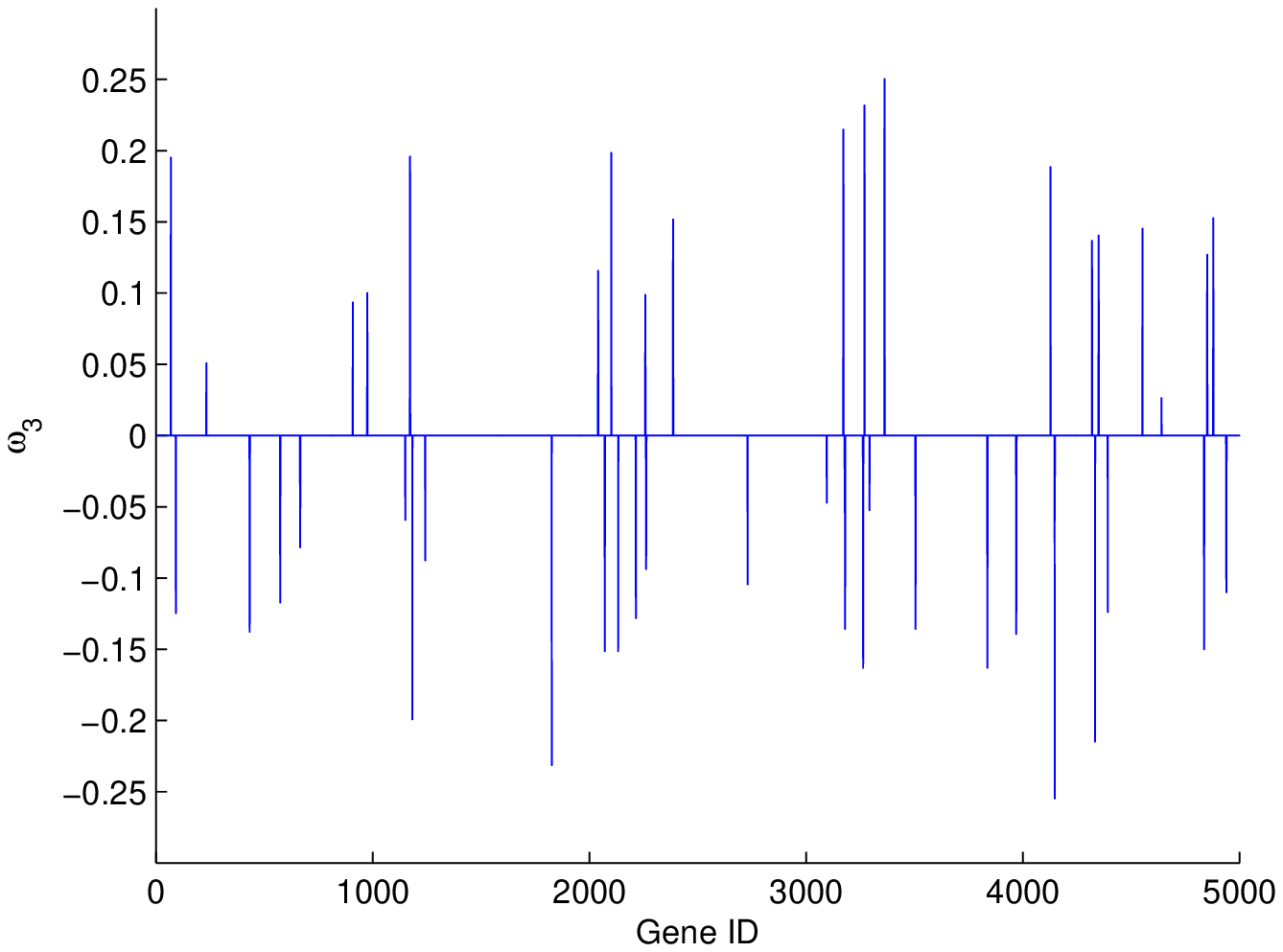}}
\end{minipage}

\begin{minipage}[h]{0.32\linewidth}
\centering (a) $\omega_1$
\end{minipage}
\begin{minipage}[h]{0.32\linewidth}
\centering (b) $\omega_2$
\end{minipage}
\begin{minipage}[h]{0.32\linewidth}
\centering (c) $\omega_3$
\end{minipage}

\caption{\label{Figure-Omega}$\omega_j$, $j=1, 2, 3$ generated by POCRE for SCD1.}
\end{figure}

With each of EN, lasso, and POCRE, we separately build up regression models for each of the three physiological phenotypic values. The results are presented in Table~\ref{Table-RealData}. Overall, lasso tends to select small number of predictors, and also reports the largest SSPE. On the other hand, POCRE reports the smallest SSPE for each phenotype, and selects smaller number of predictors for both SCD1 and PEPCK, but larger number of predictors for GPAT than EN. POCRE generates three components for SCD1 (see Figure~\ref{Figure-Omega}), and one component for each of the other two phenotypes (results not shown).

\begin{table}
\caption{\label{Table-RealData} Summary on Real Data Analysis}
\centering
\begin{tabular}{c | c c c c c c c}
\hline\hline
& \multicolumn{4}{c}{\underline{Sum of Squared Prediction Error}} & \multicolumn{3}{c}{\underline{Number of Selected Genes}}\\
Method & SCD1 & GPAT & PEPCK & Total & SCD1 & GPAT & PEPCK \\ \hline
EN & 3.96 & 22.07 & 2.59 & 28.62 & 255 & 34 & 5000 \\
Lasso & 6.38 & 22.07 & 2.87 & 31.32 & 1 & 34 & 8 \\
POCRE & 3.15 & 16.11 & 1.93 & 21.19 & 195 & 106 & 58 \\
\hline
\end{tabular}
\end{table}


We also fit a multivariate-response regression model for the three phenotypes using POCRE. Four common components are generated using a total of 277 genes. The resultant model reports SSPE for a total of 22.85 (i.e., 2.79, 18.14, and 1.93 for SCD1, GPAT, and PEPCK, respectively). The two regression models built by POCRE share only 21 genes for SCD1, 59 genes for GPAT, and 36 genes for PEPCK, although they report similar SSPE values.


\section{Discussion}

Effective dimension reduction is crucial for a successful analysis of $p\gg n$ data. Traditional unsupervised dimension reduction can be used to exclude many features from constructed sparse predictors, but the false discovery rate (FDR) can be very high. On the other hand, available supervised dimension reduction, such as PLS, ignores the sparse nature of the underlying signatures. Furthermore, all these methods assume that the predictors are accurately measured, and do not incorporate functional relatedness of candidates. As a result, despite years of searching, only a handful of predictive biomarkers have advanced to general clinical practice. Clearly, more effective approaches are called if the true potential of predictive molecular signatures is to be realized.



POCRE builds up orthogonal components by aggregating contribution of predictors along the direction which maximizes their correlations to the response variables or residuals (when predictors are standardized). It sequentially constructs these orthogonal components by finding penalized leading principal components. The involved computation is efficient and feasible for large $p$ small $n$ data. As in Section 7 which presented a training dataset with $n=50$ and $p=22,690$, POCRE, coded in MATLAB$^\mathrm{\scriptsize \textregistered}$, took less than two minutes to fit the regression model with four components (the tuning parameter was set at $\lambda=0.75$, and it was run on a desktop computer with Intel$^\mathrm{\scriptsize \textregistered}$ 3.0GHz Core$^{\mathrm{\scriptsize TM}}$ 2 Duo CPU).

POCRE implements the penalization via an empirical Bayes thresholding. Since this empirical Bayes thresholding is constructed with a sparsity-adaptive prior, POCRE is automatically enabled to select sparse variables in the large $p$ small $n$ paradigm. As shown in the simulation studies, it provides a clear and significant benefit to the general task of variable selection in the large $p$ small $n$ paradigm, even with clustered predictors or noisy predictors. It confirmed the utility of the new method in molecular profiling, thus indicating an enormous promise for its use in transcriptional profiling (genomics), protein profiling (proteomics), methylation profiling (epigenomics), and metabolite profiling (metabolomics). The full potential of the new framework, however, lies in providing breakthrough solutions to implementing the Bayesian penalization for structured noisy features.


\section*{Acknowledgements}

The authors thank Jayanta K. Ghosh for his helpful comments.

\section*{Appendix A: Proof of theorem 1}

Since for each $j$, $cov(\tilde{X}_{j+1}, \omega_j^T \tilde{X}_j) = 0$, then for any $l>0$,
\[
cov(\omega_{j+l}^T\tilde{X}_{j+l}, \omega_j^T\tilde{X}_j) = \omega_{j+l}^T \left\{\prod_{m=1}^{l-1}(I-\theta_{j+l-m}\omega_{j+l-m}^T)\right\} cov(\tilde{X}_{j+1}, \omega_j^T\tilde{X}_j) = 0,
\]
which proves that $\varpi_{1}^T X$, $\varpi_{2}^T X$, $\cdots$, are uncorrelated and therefore orthogonal.

On the other hand,
\[
\tilde{Y}_{l+1} = \tilde{Y}_l - \vartheta_l \varpi_l^T X =\cdots = Y - \sum_{j=1}^l \vartheta_j\varpi_j^T X.
\]
Suppose $\tilde{Y}_{l+1}$ is uncorrelated to $\tilde{X}_{l+1}$. Then,
\[
E[Y|X] = \sum_{j=1}^l \vartheta_j\varpi_j^T X + E[\tilde{Y}_{l+1}|X]
\]

Note that
\[
\tilde{X}_{l+1} = \tilde{X}_{l} - \theta_{l} \omega_{l}^T \tilde{X}_{l} = \cdots = X - \sum_{j=1}^{l}\theta_{j} \omega_{j}^T \tilde{X}_{j} \Longrightarrow X = \tilde{X}_{l+1} + \sum_{j=1}^{l}\theta_{j} \omega_{j}^T \tilde{X}_{j}.
\]
Therefore,
\[
cov(\tilde{Y}_{l+1}, X) = cov(\tilde{Y}_{l+1}, \tilde{X}_{l+1}) + \sum_{j=1}^{l}cov(\tilde{Y}_{l+1}, \omega_{j}^T \tilde{X}_{j})\theta_{j}^T = 0.
\]
Denote $\tilde{Y}_{l+1} = \tilde{\beta}^TX+\epsilon$, then
\[
\tilde{\beta}^T V = 0 \Longrightarrow \tilde{\beta}^TV\tilde{\beta} = cov(\tilde{\beta}^TX, \tilde{\beta}^TX ) = 0 \Longrightarrow \tilde{\beta}^TX = 0,
\]
which implies that $E[\tilde{Y}_{l+1}|X]=0$, and concludes the proof.

\section*{Appendix B: Proof of theorem 3}


Denote
\begin{eqnarray*}
(\hat{\alpha}(\kappa),\tilde{\gamma}(\kappa)) = argmin_{\alpha,\gamma: \|\alpha\|=1} \left\{ \left\|\mathbf{M}-\mathbf{M}\frac{\gamma}{1+\kappa}\alpha^T\right\|^2 + \kappa \left\|\frac{\gamma}{1+\kappa}\right\|^2 + p_{\lambda}\left(\frac{\gamma}{1+\kappa}\right) \right\}.
\end{eqnarray*}
Then $\tilde{\gamma}(\kappa)/\|\tilde{\gamma}(\kappa)\|=\hat{\gamma}(\kappa)/\|\hat{\gamma}(\kappa)\|$.

Since
\begin{eqnarray*}
\lefteqn{\left\|\mathbf{M}-\mathbf{M}\frac{\gamma}{1+\kappa}\alpha^T\right\|^2 + \kappa \left\|\frac{\gamma}{1+\kappa}\right\|^2 + p_{\lambda}\left(\frac{\gamma}{1+\kappa}\right)} \\
&= & tr(\mathbf{M}^T \mathbf{M}) + \frac{1}{1+\kappa}\left\{-2\gamma^T \mathbf{M}^T \mathbf{M} \alpha + \frac{1}{1+\kappa}tr\left(\alpha\gamma^T \mathbf{M}^T \mathbf{M}\gamma\alpha^T \right) +\frac{\kappa}{1+\kappa}\gamma^T\gamma+p_{\lambda}(\gamma)\right\} \\
& = & tr(\mathbf{M}^T \mathbf{M}) + \frac{1}{1+\kappa}\left\{-2\gamma^T \mathbf{M}^T \mathbf{M} \alpha + \gamma^T \frac{\mathbf{M}^T \mathbf{M}+\kappa I}{1+\kappa}\gamma+p_{\lambda}(\gamma)\right\}.
\end{eqnarray*}
Therefore,
\[
(\hat{\alpha}(\kappa),\tilde{\gamma}(\kappa)) = argmin_{\alpha,\gamma: \|\alpha\|=1} \left\{-2\gamma^T \mathbf{M}^T \mathbf{M} \alpha + \gamma^T \frac{\mathbf{M}^T \mathbf{M}+\kappa I}{1+\kappa}\gamma+p_{\lambda}(\gamma) \right\},
\]
which implies
\[
(\hat{\alpha}(\infty),\tilde{\gamma}(\infty)) = argmin_{\alpha,\gamma: \|\alpha\|=1} \left\{-2\gamma^T \mathbf{M}^T \mathbf{M} \alpha + \|\gamma\|^2+p_{\lambda}(\gamma) \right\} = (\hat{\alpha},\hat{\gamma}).
\]

\end{document}